\documentclass[prl,twocolumn,showpacs,superscriptaddress,floatfix,footinbib,amsmath,amssymb]{revtex4}

\usepackage[utf8]{inputenc}
\usepackage{graphicx}
\usepackage[normalem]{ulem}
\usepackage[usenames,dvipsnames]{color}
\usepackage[nointegrals]{wasysym}
\usepackage{siunitx}
\usepackage{bbm}
\usepackage{commands}

\usepackage{upgreek}
\usepackage{blkarray}
\usepackage{multirow}

\usepackage{tikz}
\newcommand{\tikzcircle}[1][black, thick, fill=red]{\tikz[baseline=-.75ex]\draw[#1,radius=2.5pt] (0,0) circle ;}%

\begin{document}


\title{Electric-dipole-induced spin resonance in a lateral double quantum dot incorporating two single domain nanomagnets}

\author{F. Forster}\affiliationMunich
\author{M. Mühlbacher}\affiliationMunich
\author{D. Schuh}\affiliationRegensburg
\author{W. Wegscheider}\affiliationRegensburg\affiliationETH
\author{S. Ludwig}\thanks{Present address: Paul-Drude-Insti\-tut f\"ur Fest\-k\"or\-per\-elek\-tro\-nik, Haus\-vog\-tei\-platz 5--7, 10117 Berlin, Germany}\affiliationMunich

\date{\today}


\begin{abstract}
On-chip magnets can be used to implement relatively large local magnetic field gradients in nanoelectronic circuits. Such field gradients provide possibilities for all-electrical control of electron spin-qubits where important coupling constants depend crucially on the detailed field distribution. We present a double quantum dot (QD) hybrid device laterally defined in a GaAs\,/\,AlGaAs heterostructure which incorporates two single domain nanomagnets. They have appreciably different coercive fields which allows us to realize four distinct configurations of the local inhomogeneous field distribution. We perform dc transport spectroscopy in the Pauli-spin blockade regime as well as electric-dipole-induced spin resonance (EDSR) measurements to explore our hybrid nanodevice. Characterizing the two nanomagnets we find excellent agreement with numerical simulations. By comparing the EDSR measurements with a second double QD incorporating just one nanomagnet we reveal an important advantage of having one magnet per QD: It facilitates strong field gradients in each QD and allows to control the electron spins individually for instance in an EDSR experiment. With just one single domain nanomagnet and common QD geometries EDSR can likely be performed only in one QD.

\end{abstract}

\maketitle

\section{introduction}

At cryogenic temperatures semiconductor based quantum dots (QD) can be used to create well defined quantum states of arbitrarily few localized electrons. The electron spins of these states provide a playground for exploring quantum mechanics in an interacting solid state environment and are heavily studied for possible applications in quantum information processing \cite{Loss1998,Petta2005,Koppens2005,Taylor2005b,Tokura2006,Koppens2006,Laird2007,Nowack2007,Hanson2007,Ladriere2008,Nadj-Perge2010,Obata2010,Shin2010b,Bluhm2011,Gaudreau2012,Maune2012,Hao2014,Kawakami2014,Yoneda2014} The coherent dynamics of electron spins can be accessed in an electron spin resonance (ESR) experiment. To control a QD based spin qubit on a time scale shorter than its dephasing time such an ESR experiment would require a magnetic field modulated at radio frequencies (rf) with an amplitude of few millitesla. Combining such a large rf modulation to an (externally applied) macroscopic magnetic field with cryogenic temperatures of $T\ll1\,$K, required for long spin lifetimes, is a major technical challenge. Obstacles are oscillating strong mechanical forces between macroscopic perpendicular magnets and eddy currents caused by induction, both causing severe heating and mechanical oscillations. To overcome these problems, on-chip methods to locally manipulate electron spins have been developed. A first breakthrough in locally controlling QD based spin qubits was based on the exchange coupling between two electrons located in adjacent tunnel coupled QDs \cite{Petta2005}. This all-electrical method makes use of the direct dependence of the singlet-triplet splitting on gate voltages, while the latter can be rf modulated in a straightforward way \cite{Petta2005,Taylor2005b,Bluhm2011,Maune2012,Gaudreau2012,Hao2014}. Because the exchange coupling between two electrons is subject to fluctuations of the local potential, it is, however, desirable to be able to manipulate the spin of a single electron localized in a QD as well. In a standard ESR approach this is, in principle, possible with an on-chip magnetic antenna \cite{Taylor2005b,Koppens2006}. This approach has the disadvantage of needing a relatively strong current through an on-chip micro-wire which causes parasitic heating of the sample. The capacitive coupling between the strongly driven antenna and the QD leads can furthermore cause electron pumping via an unwanted modulation of the leads chemical potentials. Alternative methods are based on electric-dipole-induced spin resonance (EDSR) where a periodic spatial motion of an electron gives rise to an oscillating (effective) magnetic field, an rf driving force.
The rf spatial oscillation of an electron is thereby induced by modulating the voltage of one of the metal gates defining the QD. It has been demonstrated that the necessary inhomogeneous effective magnetic field can be provided by the spin-orbit interaction \cite{Nowack2007,Nadj-Perge2010} or even the spatial fluctuations of the hyperfine interaction between the electron and many nuclei \cite{Laird2007}. Unfortunately, both these interactions also promote dephasing of the qubit; the spin-orbit interaction via coupling electrons and phonons, while the hyperfine interaction couples the electron spin dynamics to the thermal fluctuations of nuclear spins \cite{Cywinski2009a, Bluhm2010}. Consequently, it would be beneficial for spin qubit applications to use materials combining a small spin-orbit interaction with no nuclear spins,  e.g.\ $^{28}$Si and $^{12}$C. However, this would require another mechanism to facilitate EDSR.

An elegant option employs the inhomogeneous stray field near the edge of an on-chip magnet. A spatial oscillation of an electron localized in such an inhomogeneous field then directly translates into a modulation of the magnetic field. In past experiments, relatively wide (width of $\sim\upmu$m) on-chip magnets in the vicinity of double QDs have been used in order to create a strong field gradient \cite{Tokura2006,Laird2007,Ladriere2008,Obata2010,Shin2010b,Kawakami2014,Yoneda2014}. The disadvantage of such a large magnet are its multiple magnetic domains at zero external magnetic field, $\Bext$, which lead to a small and rather uncontrolled stray field. A sizable $\Bext$ on the order of a Tesla is then needed to align the domains and thereby create a strong inhomogeneous magnetic field at the QD. Multiple domains can be avoided by reducing the on-chip magnet's lateral dimensions until its shape anisotropy yields a single-domain groundstate. In a previous project, we have already realized an on-chip single-domain nanomagnet. It yields a sizable inhomogeneous stray field $\Bm$ independent of $\Bext$ and, therefore, provides interesting possibilities for nanoelectronic circuits, in particular at $\Bext\simeq0$. As an example we have demonstrated that this new regime can be utilized for very efficient hyperfine induced nuclear spin manipulation and have indeed reached much stronger nuclear spin polarizations than previously reported for lateral QDs \cite{Petersen2013}.

\begin{figure}
\includegraphics[width=1\columnwidth]{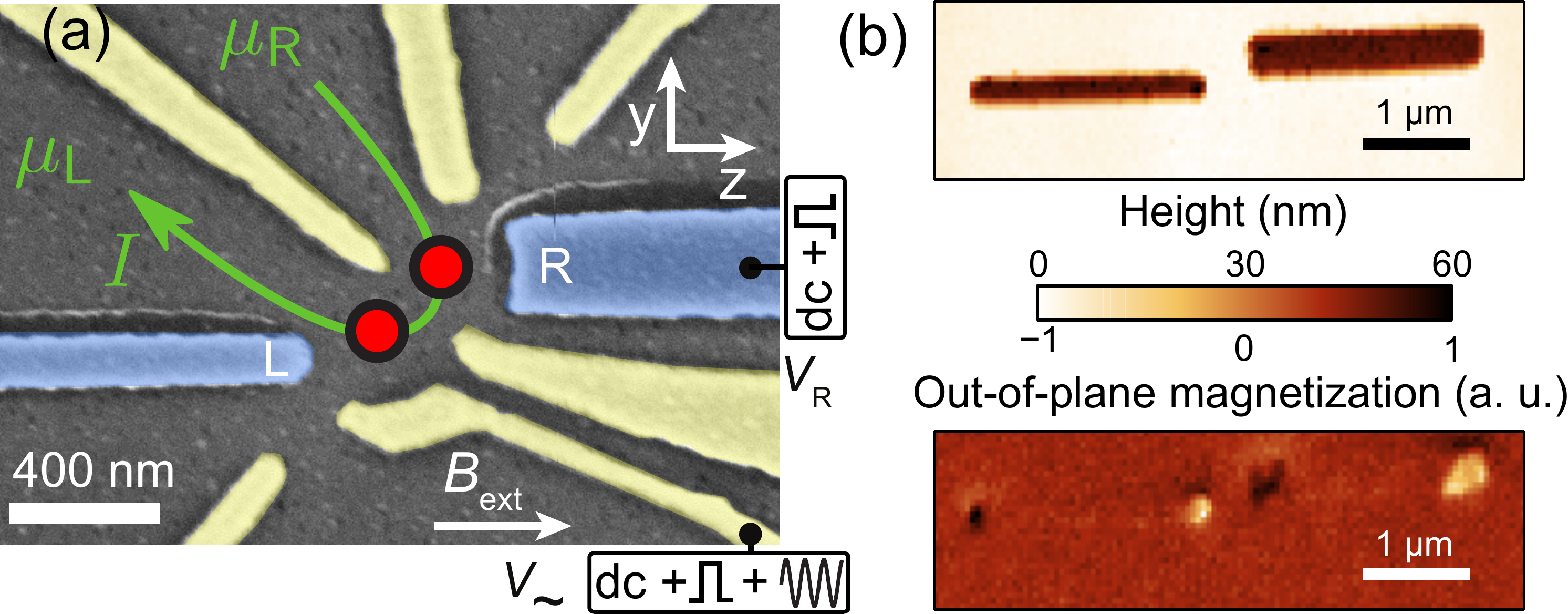}
\caption{Sample layout: (a) Scanning electron microscope image of the wafer surface. The GaAs surface is dark gray, gold gates are shown in yellow, magnetic cobalt gates  in blue. Both magnets are $\simeq 2\,\upmu$m long and $\simeq60\,$nm high, the left one (L) is $\simeq 100\,$nm and the right one (R) $\simeq 230\,$nm wide. Red filled circles indicate possible QD positions in the two-dimensional electron system 85\,nm beneath the surface; actual positions depend on gate voltages and disorder potential. The voltages applied to gates R and $\sim$ are radio frequency modulated for EDSR measurements. (b) Magnetic force microscopy measurement of the magnets before the gold gates were processed; height profile in the upper panel and out of plane magnetization in the lower panel indicating single domain magnetization of both magnets.
}
\label{fig:sample}
\end{figure}
Here we present an innovative double QD hybrid design which incorporates two single domain nanomagnets. We replaced two of the usual gold gates with ferromagnetic cobalt gates (\fig{fig:sample}{a}). At small $\Bext$, the two magnets can be magnetized in a parallel (as in \fig{fig:sample}{b}) or antiparallel configuration, giving rise to two very different inhomogeneous magnetic field distributions, an interesting possibility for spintronics applications. The double QD is defined in the two-dimensional electron system (sheet density: $1.19 \times 10^{11}\, \text{cm}^{-2}$, mobility: $0.36 \times 10^6\,\text{cm}^2 /\text{Vs}$) of a GaAs/AlGaAs heterostructure 85\,nm beneath its surface (\fig{fig:sample}{a}). 
We have prepared the double QD in the two-electron Pauli-spin blockade regime with one electron in each dot, in order to employ spin-to-charge conversion.  
To determine static properties such as the coercive fields of the two magnets we have used dc measurements and have explored the electron spin dynamics with EDSR measurements. 

Depending on the double QD geometry we have found either one or two electron-spin resonances. Two resonances corresponding to different $\Bm$ in the two dots would allow to study coupled spin qubits in a double QD. However, two resonances can only be resolved under three conditions:  
(i) a sizable magnetic field difference between the dots, (ii) a sufficiently large magnetic field gradient in each dot and (iii) a strong enough capacitive coupling between each dot and an rf-driven gate. While (iii) is straightforward to fulfill, in this article we demonstrate that the remaining conditions (i) and (ii) can be met by employing two single domain nanomagnets. As we merely replace gold gates by magnetic cobalt gates our scenario can be scaled up to multi qubit systems. Because single domain magnets are also useful at $\Bext=0$ they allow spintronics experiments beyond the scope of previous experiments with only one (usually multidomain) on-chip magnet.

\section{experimental setup}

\begin{figure*}
\includegraphics[width=2\columnwidth]{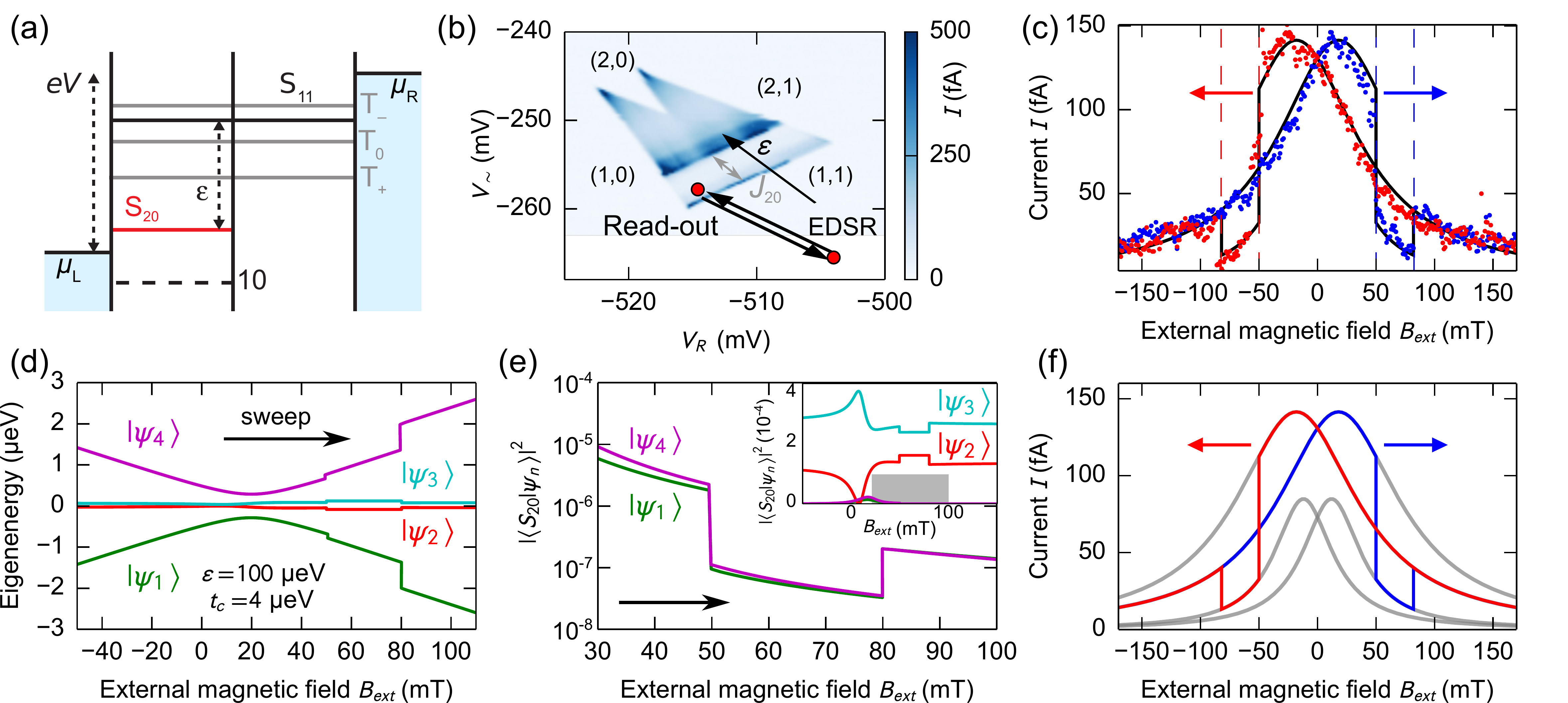}
\caption{ 
(a) Sketch showing the relevant two-electron levels of the double QD in Pauli-spin blockade and the lead chemical potentials ($\mu_\text R-\mu_\text L=1\,$meV). All horizontal lines depict chemical potentials, vertical lines indicate tunnel barriers, blue areas correspond to occupied states of the degenerate Fermi liquid in the leads. 
(b) Corresponding charge stability diagram at $\Bext=0$, presenting the current through the double QD. The numbers of electrons charging the left, right dot in stable areas of the stability diagram are indicated in parentheses. For our EDSR measurements the system was pulsed between the two red dots; details in main text.
(c) Current at $\epsilon \simeq 100\,\upmu$eV as function of $\Bext$ which was swept at a rate of $30$ mT$/$min. Arrows indicate the sweep directions.
(d) Numerically calculated eigenenergies  of the $(1,1)$ states of the system Hamiltonian \eqref{eq:hamiltonian} as a function of $\Bext$. $\left|\psi_{1,4}\right>$ are almost identical to \Tpm\ and $\left|\psi_{2,3}\right>$ are superpositions of \Tzero\ and \Slr. For the used parameters, $\epsilon=100\,\upmu$eV and $\tc=4\,\upmu$eV, the (2,0) state has a much lower energy. Vertical steps at $\Bext\simeq50,\,80\,$mT indicate discontinuities of \Bm\ caused by the numerically included switching of the nanomagnets. (e) Overlap of the $(1,1)$ eigenstates shown in d with the \Sll\ state (which would be zero for the \Tlr\ states in a homogeneous field): all four states in the inset. The main panel is a magnified view of the gray area. It contains the two states with the smallest overlap, the current limiting bottleneck states (almost \Tpm). (f) Sketch of the four current resonances belonging to the four different magnets' configurations. The red and blue line resemble the switching behaviour of the actual current in (c) (arrows indicate sweep directions).}
\label{fig:stability}
\end{figure*}

Our measurements probe the dc current $I$ (green arrow in \fig{fig:sample}{a}) which passes through the double QD in response to a constant voltage $V=(\mu_\text R-\mu_\text L)/e=1\,$mV applied across it. Figure \ref{fig:stability}a
illustrates the double QD configuration by sketching the chemical potentials $\mu_\text{L,R}$ of the leads and those of the relevant double QD states as horizontal lines, while vertical lines indicate tunnel barriers. We denote  $(n,m)$ the charge configuration of the double QD with $n$ electrons in the left and $m$ in the right dot and consider the single electron charge transfer characterized by the following tunneling cycle:  $(1,0)\to(1,1)\to(2,0)\to(1,0)$, where the transition $(1,1)\to(2,0)$ constitutes a bottleneck: Both configurations, $(1,1)$ and $(2,0)$, are composed of three triplets, collectively denoted by \Tlr\ and \Tll, and one singlet, \Slr\ and \Sll. We define the detuning $\epsilon$ as the energy difference between \Slr\ and \Sll. For $\epsilon<0$ tunneling processes $(1,1)\to(2,0)$ are blocked by energy conservation (Coulomb blockade) and $I$ is close to zero. The exchange splitting between singlets and triplets is much higher if two electrons are in the same dot, $\Jll\gg\Jlr$. Hence, at $\epsilon\simeq0$ the \Tll\ states are highly elevated compared to \Tlr. As a consequence, for $0<\epsilon<\Jll$ the transition $\Tll\to\Slr$ is forbidden by the Pauli principle (Pauli-spin blockade), as long as \Slr\ and \Tlr\ remain decoupled.
In our case the inhomogeneous $\Bm$ mixes \Slr\ and \Tlr\ states near where their eigenenergies are equal \cite{Petersen2013}. 
In the stability diagram plotted in \fig{fig:stability}{b} this coupling gives rise to a narrow stripe of $|I|>0$ near $\epsilon=0$. We stress that the hyperfine interaction\, which also couples \Slr\ and \Tlr, is only a weak perturbation compared to the effect of $\Bm$. For $\epsilon\ge\Jll$ the transition $\Tlr\to\Tll$ lifts the Pauli-spin blockade and a sizable current flows (in our double QD $\Jll\simeq300\,\upmu$eV). The tiny but non-vanishing current visible in \fig{fig:stability}{b} for $0<\epsilon<\Jll$ is dominantly caused by higher order processes such as the co-tunneling $\Tlr\to\Tll\to(1,0)$, where \Tll\ is energetically forbidden.

In the stability diagram in \fig{fig:stability}{b}, the current carrying region corresponding to the absence of Coulomb blockade is composed of two overlapping triangles. Above we have described the tunneling cycle which gives rise to the lower left triangle; the second triangle corresponds to an alternative cycle $(2,1)\to(1,1)\to(2,0)\to(2,1)$. Nevertheless, the above explanations apply to both cycles, as they are both bottlenecked by the same transition $(1,1)\to(2,0)$.

In previous devices on-chip magnets were separated from the heterostructure by a layer of metal gates and, with the exception of Ref.\ \onlinecite{Petersen2013}, in addition by a second electrically isolating layer. Here, we simplify the structure and bring the magnets closer to the QDs by replacing two gold-gates (yellow in \fig{fig:sample}{a}), used to define the double QD, by ferromagnetic cobalt-gates (blue). 
Based on simulations with OOMMF \cite{Donahue1999} we have tailored the stray fields of the magnets and have optimized their geometries and positions to maximize stray field and field gradient between the two dots on the one hand and to guarantee full tunability of the double QD by applying gate voltages on the other hand. The advantage of using two instead of just one magnet is twofold: first, two magnets can be positioned to provide strongly inhomogeneous and different magnetic fields in two adjacent dots (conditions (i) and (ii) above) which facilitates EDSR measurements in both dots. Second, at moderate $\Bext$ two separate magnets allow for two very different stray field distributions across the double QD corresponding to either parallel or anti-parallel magnetization of the two magnets (see \fig{fig:OOMMF}{}). The magnetization of each nanomagnet can thereby be reversed by sweeping $\Bext$ beyond its coercive field and antiparallel to its present magnetization. The different width of the two magnets gives rise to individual coercive fields. 
Consequently, we can choose between parallel and anti-parallel magnetization at relatively small $\Bext$.  

\section{system Hamiltonian}

To model the dynamics of our double QD we assume that an electron localized in the left respective right dot experiences the local magnetic field $\mathbf{B}^\text{L,R} = \Bext + \Bm^\text{L,R}$. We thereby neglect the hyperfine interaction between the electron and nuclear spins, the spin-orbit interaction and the exchange interaction which in our case are all small perturbations compared to the coupling induced by the inhomogeneous $\Bm$. For simplicity we define the average field in the two dots $\Bs = \left( \Bl + \Br \right)/2$, their difference field $\Bd =  \Bl - \Br$ and the field operators $\overline{B}_\pm = \overline{B}_x \pm i \overline{B}_y$, $\Delta B_\pm = \left(\Delta B_x \pm i \Delta B_y\right)/2$ akin to spin raising and lowering operators. With the quantization axis $\hat{z}$ defined parallel to $\Bext$, the matrix
representation of the (semiclassical) total Hamiltonian in the basis spanned
by the diabatic singlet and triplet states \{\Tp, \Tzero, \Tm, \Slr, \Sll \}
is then

\begin{widetext}
\parbox{.9\linewidth}{
\begin{center}
 \raisebox{-1.2em}{$\displaystyle H = g\mB\;$}
\begin{blockarray}{ccccc@{\hspace*{13pt}}|@{\hspace*{5pt}}c}
\hspace{5ex}\Tp\hspace{5ex} & \hspace{5ex}\Tzero\hspace{5ex} & \hspace{5ex}\Tm\hspace{5ex} & \hspace{5ex}\Slr\hspace{5ex} & \hspace{5ex}\Sll\hspace{3ex} &\\\cline{1-6}
&&&&&\\
 \begin{block}{(ccccc)@{\hspace*{20pt}}|@{\hspace*{5pt}}l}
 $\displaystyle \overline{B}_z$ & $\overline{B}_-/\sqrt{2}$ & $0$ & $-\Delta B_-/\sqrt{2}$ & $0$ & $\Tp = \ket{\uu}$ \\[0.7em]
 $\overline{B}_+/\sqrt{2}$ & $0$ & $\overline{B}_-/\sqrt{2}$ & $\Delta B_z/2$ & $0$ & $\Tzero = \left( \ket{\ud} + \ket{\du} \right)/\sqrt{2}$ \\[0.7em]
 $0$ & $\overline{B}_+/\sqrt{2}$ & $- \overline{B}_z$ & $ \Delta B_+/\sqrt{2}$ & $0$ & $\Tm  = \ket{\dd}$ \\[0.7em]
 $-\Delta B_+/\sqrt{2}$ & $ \Delta B_z/2$ & $\Delta B_-/\sqrt{2}$ & $0$ & $\tc/2g\mB$ & $\Slr = \left( \ket{\ud} - \ket{\du} \right)/\sqrt{2}$ \\[0.7em]
 $0$ & $0$ & $0$ & $\tc/2g\mB$ & $-\epsilon/g\mB$ & $\Sll = \ket{0,\ud}$ \\[0.3em]
 \end{block}
\end{blockarray}
\end{center}} \hfill\parbox{0pt}{
\begin{equation}
  \label{eq:hamiltonian}
\end{equation}}
\end{widetext}
\noindent
where $\tc$ denotes the interdot tunnel coupling between the two dots. The matrix representation (\ref{eq:hamiltonian}) illustrates that the $x$- and $y$-components of the difference field, $\Delta B_\pm$, mix $T_\pm$ with $S_{11}$, while the $z$-component $\Delta B_z$ mixes \Tzero\ (which has no spin component along the $z$-axis). 
The average field $\Bs$ yields the Zeeman splitting of the spin-up versus spin-down states. Note that the off-diagonal terms $\overline B_\pm$, which mix $T_\pm$ with $T_0$, vanish if the quantization axis is chosen parallel to \Bs\ instead of \Bext. 

Hyperfine and spin-orbit interaction would both contribute to various matrix elements including the singlet-triplet coupling constants. In our case, however, the latter are far dominated by the time independent difference field (and we formally neglect the former contributions). In this way the nanomagnets provide a stabilization mechanism for appropriate qubit implementations which could increase the qubit coherence time in spite of the presence of nuclear spins or spin-orbit interaction.
To fully determine our double QD hybrid system we need to know the nanomagnet's strayfield as a function of $\Bext$ at the position of the two dots, the interdot tunnel coupling $\tc$ and the Lande g-factor inside the dots $g$. In the following we will employ dc current measurements and EDSR experiments to achieve this goal.

\section{Direct current measurements} \label{sec:dc-current}

To experimentally determine the coercive fields of our nanomagnets we have measured the leakage current $I$ through the spin-blockaded double QD while slowly sweeping $\Bext$ at constant detuning, $\epsilon\simeq 100\,\upmu$eV. The current at this configuration is sensitive to the mixing of the singlet and triplet states \cite{Koppens2005} and can be used to detect changes of the magnetic field differences between the QDs \cite{Petersen2013}.

In such a sweep experiment the current $I(\Bext)$ might be influenced by dynamic nuclear spin polarization (DNSP)  which can give rise to hysteresis as a function of the sweep direction of $\Bext$\cite{Ono2004,Petersen2013, Kobayashi2011}. However, it is also possible to avoid DNSP effects by preparing a fixed point at very weak polarization \cite{Petersen2013}. Here and in the EDSR experiments discussed below, DNSP is negligible and the apparent hysteresis of the measured $I(\Bext)$ visible in \fig{fig:stability}{c} has a different reason: it is related to the four distinct configurations of the magnets, each of which can be magnetized parallel or anti-parallel to $\Bext$.

In \fig{fig:stability}{c} we present $I(B_\text{ext})$ for two sweeps in opposite directions ($dB/dB_\text{ext}=\pm30\,\text{mT}/\text{min}$). We have started the sweeps at large $|B_\text{ext}|$ to ensure that both magnets are magnetized parallel to \Bext. The current maxima near $\Bext=0$ occur where all three \Tlr\ states are close to resonance with \Slr. This is a consequence of Pauli-spin blockade where the leakage current is governed by the singlet-triplet couplings: the \Tlr\ triplets mix with \Slr\ and because \Slr\ is tunnel coupled to the other singlet, \Sll, the \Tlr\ triplets mix also with \Sll. These mixings are strongest near the mutual resonances between \Slr\ and \Tlr\ and zero far away from the corresponding resonances. As $|B_\text{ext}|$ is increased the \Tp\ and \Tm\ triplets are more and more detuned from \Slr, their mixing with \Sll\ also decreases, their decay ($\Tlr\to\Sll$) slows down and they become the current limiting bottleneck states. Consequently, the current decreases. To illustrate this connection we numerically diagonalized the Hamiltonian in \eq{eq:hamiltonian} and plot in \fig{fig:stability}{d} the energies of the four relevant $(1,1)$ eigenstates versus $B_\text{ext}$ for a sweep from negative to positive fields. The apparent avoided crossing at $B_\text{ext}\simeq20\,$mT marks the point of minimal $\Bext+\Bm$, where $\overline{B}_z=0$.  This field coincides with the current maximum for $dB/dB_\text{ext}>0$ (blue in \fig{fig:stability}{c}), because here the \Tpm-\Sll\ mixing has its maximum. For $dB/dB_\text{ext}<0$ the nanomagnets would be magnetized in the opposite direction and the current maximum would occur at $B_\text{ext}\simeq-20\,$mT (as observed in the according measurement, red in \fig{fig:stability}{c}).

Continuing the sweep, we further increase \Bext\ beyond the respective coercive fields of the two magnets where their magnetizations reverse and become again parallel to \Bext. This change of magnetization instantly rearranges the overall magnetic field at the QDs and causes a steplike characteristic of the eigenenergies at the coercive fields. We show below that because of the direct relation between the singlet-triplet mixing and the nanomagnets' configuration, this leads to the sudden changes of the measured current observed in \fig{fig:stability}{c}, where we find coercive fields at $B_\text{ext}\simeq50\,$mT for the wider and $B_\text{ext}\simeq80\,$mT for the narrower magnet. These coercive fields are included in the numerics of \fig{fig:stability}{d}.

We remark that the observed current jumps occur very abruptly as a function of $\Bext$. This underlines that the nano magnets are single domain and the single domain switches as a whole once the coercive field is reached.

To phenomenologically explain the current $I(B_\text{ext})$ in \fig{fig:stability}{c}, we assume four distinct current maxima corresponding to the four possible configurations of our nanomagnets. By sweeping \Bext\ we can switch between these configurations which causes the actual current to jump between the four maxima at the corresponding coercive fields. This simple, yet reasonable model is displayed in \fig{fig:stability}{f}, where we plot two pairs of Lorentzians (gray) reflecting the symmetry properties of the problem. The larger maxima correspond to the parallel and the smaller ones to the anti-parallel magnets' configurations. For better comparability with our measurements we have added two colored curves which mimic the actual current jumps between the Lorentzians. The identical curves are also shown in \fig{fig:stability}{c} as black lines, where they reveal good agreement with the measured data. The two Lorentzians describing the parallel or anti-parallel magnets' configurations, respectively, are equal in amplitude and width. Most interesting is the observation that the overall current through the QD is considerably smaller if the two magnets are polarized anti-parallel to each other compared to their parallel configurations. It suggests that the anti-parallel magnetization causes a smaller singlet-triplet mixing of the bottleneck triplets \Tpm\ than the parallel configuration. From our Hamiltonian in \eq{eq:hamiltonian} we see that the coupling between the \Tpm-states and the singlet sub-space is proportional to the difference of the magnetic field component perpendicular to the quantization axis (approximately parallel to \Bext) between the two QDs. Thus it suggests that the perpendicular component of the field difference between the two QDs is smaller when the magnetization of the magnets is anti-parallel compared to the parallel configurations. To check this, we have approximately determined the location of the two QDs (as depicted in \fig{fig:OOMMF}~below) taking into account the numerically calculated \Bm\ combined with results of the EDSR measurements (discussed below), the gate voltage configuration (referred to as configuration I) and the measurement in \fig{fig:stability}{c}. This information provides the magnetic field components in our Hamiltonian in \eq{eq:hamiltonian} (see table \ref{tab:b_values}) and allows us to calculate its eigenenergies, shown in \fig{fig:stability}{d}, as well as the mixings between the four $(1,1)$ states and \Sll. The latter are visualized in the inset of \fig{fig:stability}{e} as function of $B_\text{ext}$ swept from negative towards positive fields. The main panel is a magnification showing the mixings of the two bottleneck states \Tpm\ near the current steps in \fig{fig:stability}{c}. It also shows a steplike characteristic at the coercive fields of the two magnets and is considerably reduced for the magnets in their anti-parallel configuration, namely between the two coercive fields. This strengthens the Pauli-spin blockade and explains the reduced current for the anti-parallel configuration of the nanomagnets in between the current steps) in \fig{fig:stability}{c}.
We close this section by noting that an accurate prediction of the current which depends on \Bm, \Bext, $\epsilon$ and \tc~would require a detailed density matrix calculation which goes beyond the scope of this article.

\section{Electric-dipole-induced spin resonance measurements}

Mixing between any two singlet and triplet states is strongly enhanced where their eigenenergies are nearly equal. If detuned from this resonance, it is possible to actively drive transitions between two levels in an EDSR experiment which regains the resonance condition by applying a proper rf magnetic field. We have performed our EDSR measurements in the $(1,1)$ configuration with $\epsilon\simeq-100\tc$ (lower red dot in \fig{fig:stability}{b}) where the two electrons are strongly localized in the two respective dots. Consequently, we expect to find two distinct EDSR resonances at the respective Zeeman energies in the two dots
%
\begin{equation}
hf=\left|g\mB\Blr\right|\simeq g\mB \left(B_\text{ext} + B^\text{L,R}_\text{nm}|_z\right)\,, \label{eq:resonance}
\end{equation}
where $f$ is the modulation frequency, $h$ the Planck constant and $B^\text{L,R}_\text{nm}|_z$ the z-component of $\left|\Bm^\text{L,R}\right|$. The approximation in \eq{eq:resonance} is fair for $(B_\text{ext} + B^\text{L,R}_\text{nm}|_z)^2 > B^\text{L,R}_\text{nm}|_x^2 + B^\text{L,R}_\text{nm}|_y^2$. $B_\text{ext}$ and $B^\text{L,R}_\text{nm}|_z$ can have identical or opposite signs depending on the magnets configuration. The resonance condition \eq{eq:resonance} allows us to directly probe the $g$-factor as well as the $z$-component of \Bm\ in the two dots and therefore also $\Bdz$.  

\begin{figure}
\includegraphics[width=1\columnwidth]{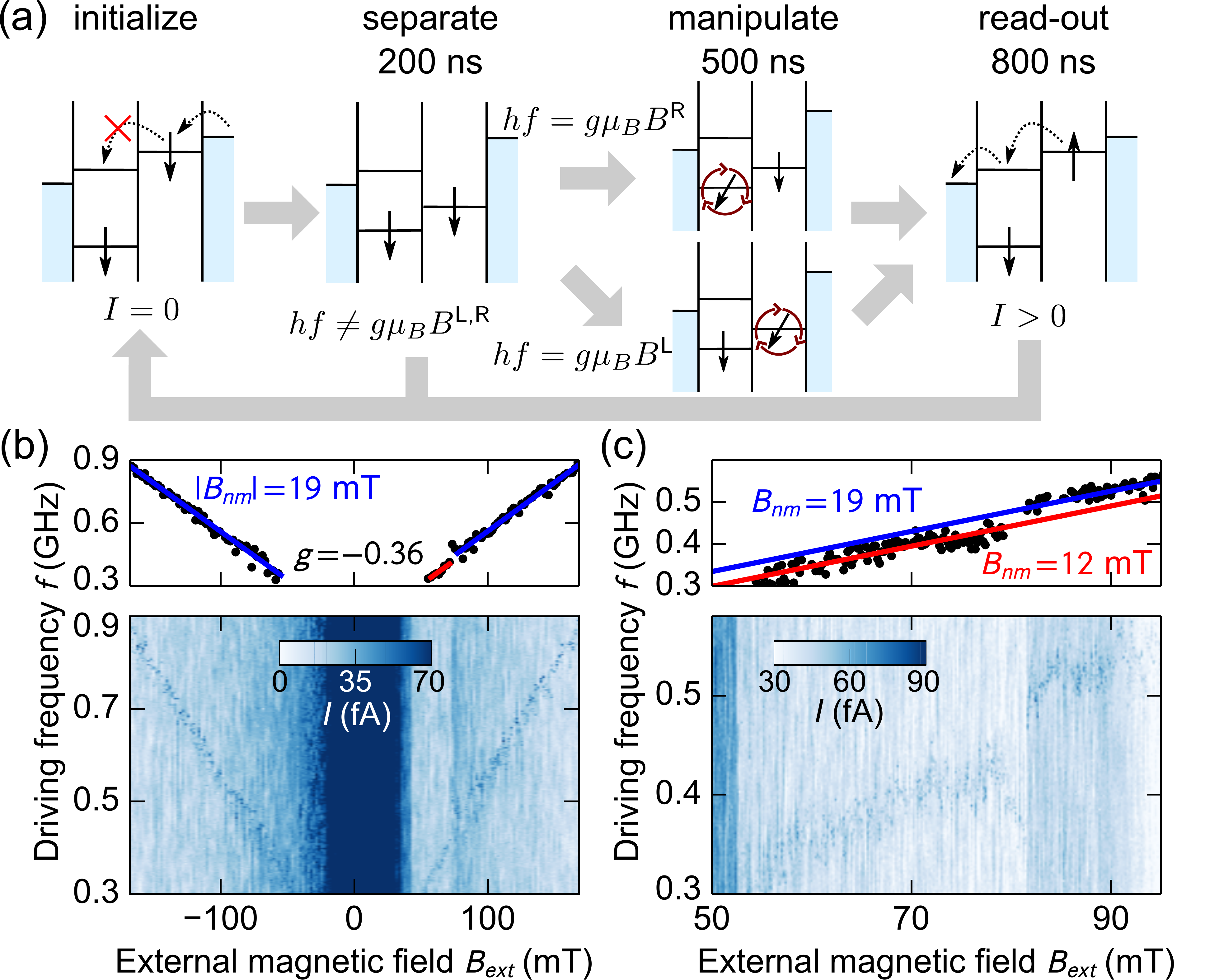}
\caption{(a) EDSR measurement scheme, from left to right: initialization in Pauli-spin blockade $\to$ separation of the QD states from the leads (Coulomb blockade) $\to$ EDSR manipulation by rf-modulation of gate voltage $\to$ read-out ($I>0$ for EDSR resonance) and re-initialization. (b) Current $I(B_\text{ext},f)$ through the DQD (bottom panel) and position of current maxima at EDSR resonance (top panel) while driving with the pulse sequence shown in (a). (c) Same as (b) but high resolution measurement near the coercive field of the left magnet $\Bc^\text{L} \sim 80\,\text{mT}$, where the resonance line forms a step.}
\label{fig:singleresesr}
\end{figure}

Our experimental EDSR sequence is sketched in \fig{fig:singleresesr}{a}.
We start at $\epsilon\gtrsim0$ in the Pauli-spin blockade (upper red dot in \fig{fig:stability}{b}), which initializes the double QD with equal probabilities in one of the bottleneck states \Tp\ or \Tm. (The other two $(1,1)$ states, \Tzero\ and \Slr, decay quickly if occupied and eventually the system stalls in \Tp\ or \Tm.)
After 800\,ns we isolate the two electrons by pulsing the double QD  deep into Coulomb blockade to $\epsilon\simeq-100\tc$ (lower red dot in \fig{fig:stability}{b}) by changing the gate voltages $V_\text R$ and $V_\sim$ (see \fig{fig:sample}{a}) within $\simeq2\,$ns. To avoid pulse transients effects we next wait 200\,ns before we modulate $V_\sim$ for 500\,ns with a sine wave which causes both electrons to oscillate in real space. Due to the inhomogeneous \Bm\ this rf modulation directly translates into oscillations of both $\Bl$ and $\Br$. Finally, we pulse back to our starting point at $\epsilon\gtrsim0$ for read out. If the rf modulation in both dots is off-resonant the double QD stays in Pauli-spin blockade and no current flows. However, if the resonance condition \eq{eq:resonance} is fulfilled for one of the two electrons during the rf modulation, the \Slr\ singlet state becomes occupied with a finite probability. As a consequence the Pauli-spin blockade is lifted during read-out. Performing a steady state measurement by periodically repeating this sequence at a frequency of $\simeq670\,$kHz we then measure a small leakage current.

Typical results of such measurements are presented in \twofigs{fig:singleresesr}{b}{c}, where the leakage current is plotted as function of $B_\text{ext}$ and $f$. At $B_\text{ext}\simeq0$ the Pauli-spin blockade is lifted even without applying an rf-modulation as already seen in \fig{fig:stability}{c} and discussed there. This effect gives rise to the broad frequency independent current maximum at $B_\text{ext}\simeq0$. The rf modulation, however, generates additional sharp but weak current maxima along straight lines, where the resonance condition in \eq{eq:resonance} is fulfilled.
Whenever \Bext\ causes one of the nanomagnets to reverse its magnetization the resonance frequency suddenly increases according to the increase of the Zeeman energy.
In \fig{fig:singleresesr}{b}, $B_\text{ext}$ was stepped from negative towards positive fields and hence the nanomagnets reverse their magnetizations at the positive fields, $B_\text{ext}\simeq50,\,80\,$mT. Figure \ref{fig:singleresesr}{c} shows a high resolution measurement of part of \fig{fig:singleresesr}{b}. It reveals one of the expected jumps in resonance frequency at $B_\text{ext}\simeq80\,$mT.  Linear fits according to \eq{eq:resonance} suggest $|g|= 0.36 \pm 0.01$ and $B^\text L_\text{nm}|_z=19\pm 1\,$mT if both magnets are magnetized parallel to $\Bext$ and $B^\text L_\text{nm}|_z=12 \pm 1\,$mT if the narrower magnet (with the larger coercive field) is aligned anti-parallel to $\Bext$.

We close the discussion of \figs{fig:singleresesr}{b}{c} with two remarks: first, the switching of the large magnet cannot be observed in this measurement because it is masked by the broad current maximum around $B_\text{ext}=0$. Second, in \figs{fig:singleresesr}{b}{c} we find only one EDSR resonance for this particular gate voltage configuration (configuration I). We will show below that it belongs to the left QD and under which conditions a second EDSR resonance can be observed. 

\begin{figure}
\includegraphics[width=1\columnwidth]{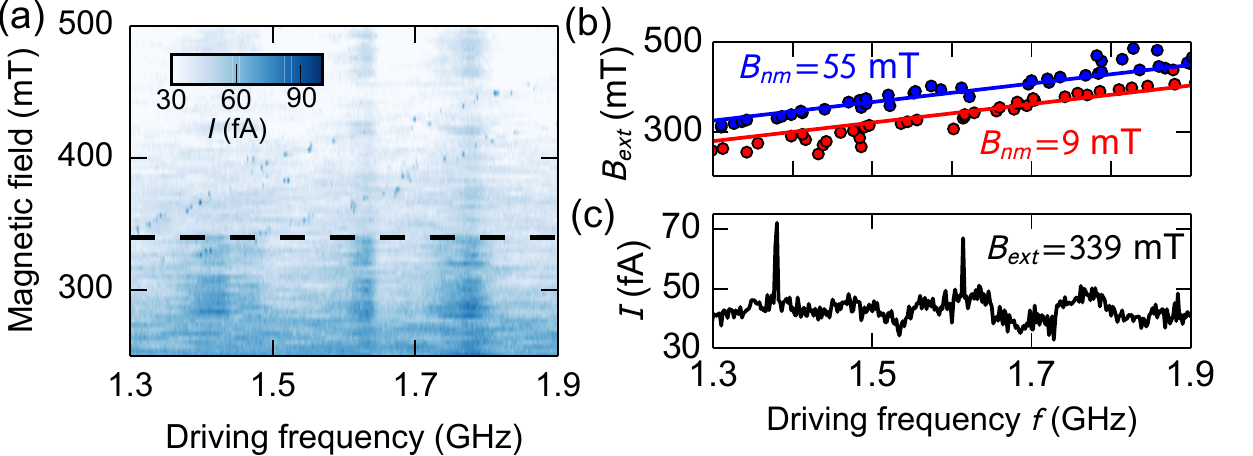}
\caption{Coexistence of two EDSR resonances: (a) $I(B_\text{ext},f)$ through the DQD as in \twofigs{fig:singleresesr}{b}{c} but for a different gate voltage configuration (configuration II). (b) Position of the current maxima at EDSR resonances in (a). Lines indicate fits with eq. \eqref{eq:resonance}. (c) Typical current trace at constant \Bext\ along dashed line in (a).}
\label{fig:doubleresesr}
\end{figure}
In \fig{fig:doubleresesr}{} we present a second EDSR measurement after retuning the double QD such that each QD is situated in close proximity to one of the two magnets. In this gate voltage configuration (II) the anti-parallel magnetization of the nanomagnets did not result in clear EDSR resonances because of strong effects of DNSP at the required weak \Bext\ values. Fortunately, the parallel magnetization of the nanomagnets allows for EDSR measurements at higher magnetic fields where DNSP is weak. In this regime we find two distinct EDSR resonances as can be seen in the raw data in \fig{fig:doubleresesr}{a} plotting $I(B_\text{ext},f)$, in panel (b) indicating the positions of sharp current maxima in (a) and also in panel (c) presenting a single frequency trace at constant \Bext. The latter contains two sharp current maxima clearly indicating two distinct EDSR resonances. Fitting \eq{eq:resonance} yields the same $|g|=0.36 \pm 0.02$ as above, but $B^\text L_\text{nm}|_z=9\pm 2\,$mT and  $B^\text R_\text{nm}|_z=55\pm 3\,$mT. The assignment of the EDSR resonances to either the left (L) or the right (R) QD is thereby based on our OOMMF simulations of \Bm.
\begin{figure}
\includegraphics[width=1\columnwidth]{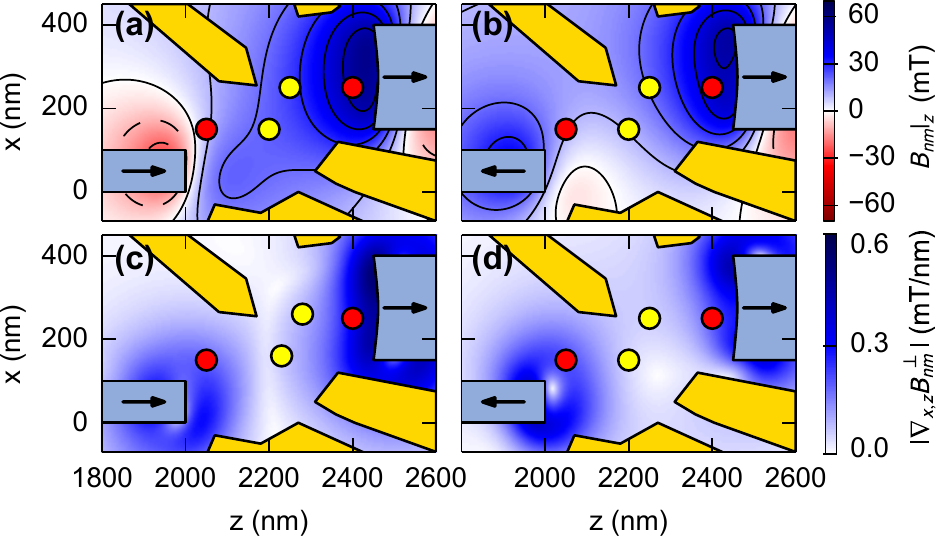}
\caption{
OOMMF simulations of \Bm\ \cite{Donahue1999}. Magnetic field component  $\Bm|_z$ along $B_\text{ext}$ (z-axis) for (a) parallel and (b) anti-parallel magnetization of the two nanomagnets. $\Bm|_z$ is relevant for the EDSR resonance condition, \eq{eq:resonance}. (c, d) Absolute value of the gradient of the perpendicular component $\Bm^\bot$ within the plane of the 2DES; $\Bm=\Bm|_z+\Bm^\bot$. The EDSR signal strength scales with $|\nabla_{x,z} \Bm^\bot|$. Colored circles in (a--d): Approximate QD center positions for the two different gate voltage tunings. Arrows indicate nanomagnets polarization directions.}
\label{fig:OOMMF}
\end{figure}
In \twofigs{fig:OOMMF}{a}{b} we plot the $\Bm|_z$ component which is relevant for the $\Tpm\leftrightarrow\Slr$ transitions considered here for the two cases, a parallel and an anti-parallel magnetization of the nanomagnets.

Comparing our numerical results with our EDSR measurements we find the approximate center coordinates of the QDs, which are marked as red circles in \figs{fig:OOMMF}{a}{d}. The used gate voltages are in agreement with these locations, where details also depend on the disorder potential. The yellow circles in \figs{fig:OOMMF}{a}{d} indicate the approximate positions of the QDs for the previous gate voltage configuration (I) discussed above. 

The strength of each EDSR resonance (i.e.\ the height of each current maximum in \fig{fig:doubleresesr}{c}) should scale with the absolute value of the local gradient of the magnetic field component perpendicular to \Bext, i.e.\ $\Bm^\bot$, along each electron oscillation path caused by the rf modulation. As we do not know the exact pathways we plot in \twofigs{fig:OOMMF}{c}{d} the absolute value of the two-dimensional derivative within the plane of the 2DES, $|\nabla_{x,z} \Bm^\bot|$. The red circles in \twofigs{fig:OOMMF}{c}{d} are clearly near derivative extrema while the situation is not so clear for the yellow circles. 
This observation provides a possible explanation for the missing second EDSR resonance in our first gate voltage configuration (yellow circles). There we can assign the observed resonance to the left QD as the simulated \Bm\ at the position of the left QD fits to the EDSR results for both magnet configurations but that of the right QD does not.
\begin{table}[b]
\centerline{configuration I (single EDSR resonance) -- $B_\text{nm}$\,(mT)}

\begin{tabular}{|c|c!{\vrule width 1pt}c|c|c|c|c|c|} 
\hline
magnetiz.\ & QD & $B_\text{nm}|_x$ & $B_\text{nm}|_y$ & $B_\text{nm}|_z$ & $\Delta B_x$ & $\Delta B_y$ & $\Delta B_z$ \\ \noalign{\hrule height 1pt}
\multirow{2}{*}{$\rightrightarrows$} & L \tikzcircle[black, thick, fill=yellow] & -15 & 6 & 17 (19) &\multirow{2}{*}{-3}  &\multirow{2}{*}{7} & \multirow{2}{*}{5}\\
 & R \tikzcircle[black, thick, fill=yellow] & -12 & -1 & 22 &  &   &\\ \hline 
\multirow{2}{*}{$\rightleftarrows$} & L \tikzcircle[black, thick, fill=yellow] & -2 & -7 & 9 (12) &\multirow{2}{*}{1} &\multirow{2}{*}{2} & \multirow{2}{*}{8}\\
 & R \tikzcircle[black, thick, fill=yellow] & -3 & -5 & 17 &  &  &  \\ \hline
\end{tabular}
\vspace{1.5ex}\
\centerline{configuration II (two EDSR resonances) -- $B_\text{nm}$\,(mT)}
\begin{tabular}{|c|c!{\vrule width 1pt}c|c|c|c|c|c|} 
\hline
magnetiz.\ & QD & $B_\text{nm}|_x$ & $B_\text{nm}|_y$ & $B_\text{nm}|_z$ & $\Delta B_x$ & $\Delta B_y$ & $\Delta B_z$ \\ \noalign{\hrule height 1pt}
\multirow{2}{*}{$\rightrightarrows$} & L \tikzcircle & 14 & 19 & 10 (9) &\multirow{2}{*}{$-1$}  &\multirow{2}{*}{7} & \multirow{2}{*}{40 (46)}\\
 & R \tikzcircle[black, thick, fill=red] & 13 & 26 & 50 (55) &  &   &\\ \hline 
\multirow{2}{*}{$\rightleftarrows$} & L \tikzcircle & -10 & -11 & 10 &\multirow{2}{*}{18} &\multirow{2}{*}{30} & \multirow{2}{*}{38}\\
 & R \tikzcircle[black, thick, fill=red] & 8 & 29 & 48 &  &  &  \\ \hline
\end{tabular}
\caption{Magnetic field components generated by the two nanomagnets (for parallel [$\rightrightarrows$] vs.\ anti-parallel [$\rightleftarrows$] magnetization) at the approximate QD positions marked in \fig{fig:OOMMF}{} by corresponding circles. Field values are calculated with OOMMF \cite{Donahue1999}; field strengths derived from measured EDSR resonances in parentheses.}
\label{tab:b_values}
\end{table}
Our numerical results at the marked QD positions are summarized and compared to our experimental findings in Table \ref{tab:b_values}.
Interestingly, for the parallel magnetization of the nanomagnets the difference fields \Bdx\ and \Bdy\ are quite weak compared to \Bdz, while \Bdx\ and \Bdy\ are also sizable for the anti-parallel magnetization. This implies that the dynamics of the \Tzero\ state is quite different for the two cases, as the coupling between \Tzero\ and \Slr\ is mediated by  \Bdx\ and \Bdy. For instance, anti-parallel magnetization would be the better choice for defining a qubit based on the states \Tzero\ and \Slr, while the parallel magnetization would be a good choice for a qubit based on \Tpm\ and \Slr.

\section{Samples with one nano magnet}
\begin{figure}
\includegraphics[width=1\columnwidth]{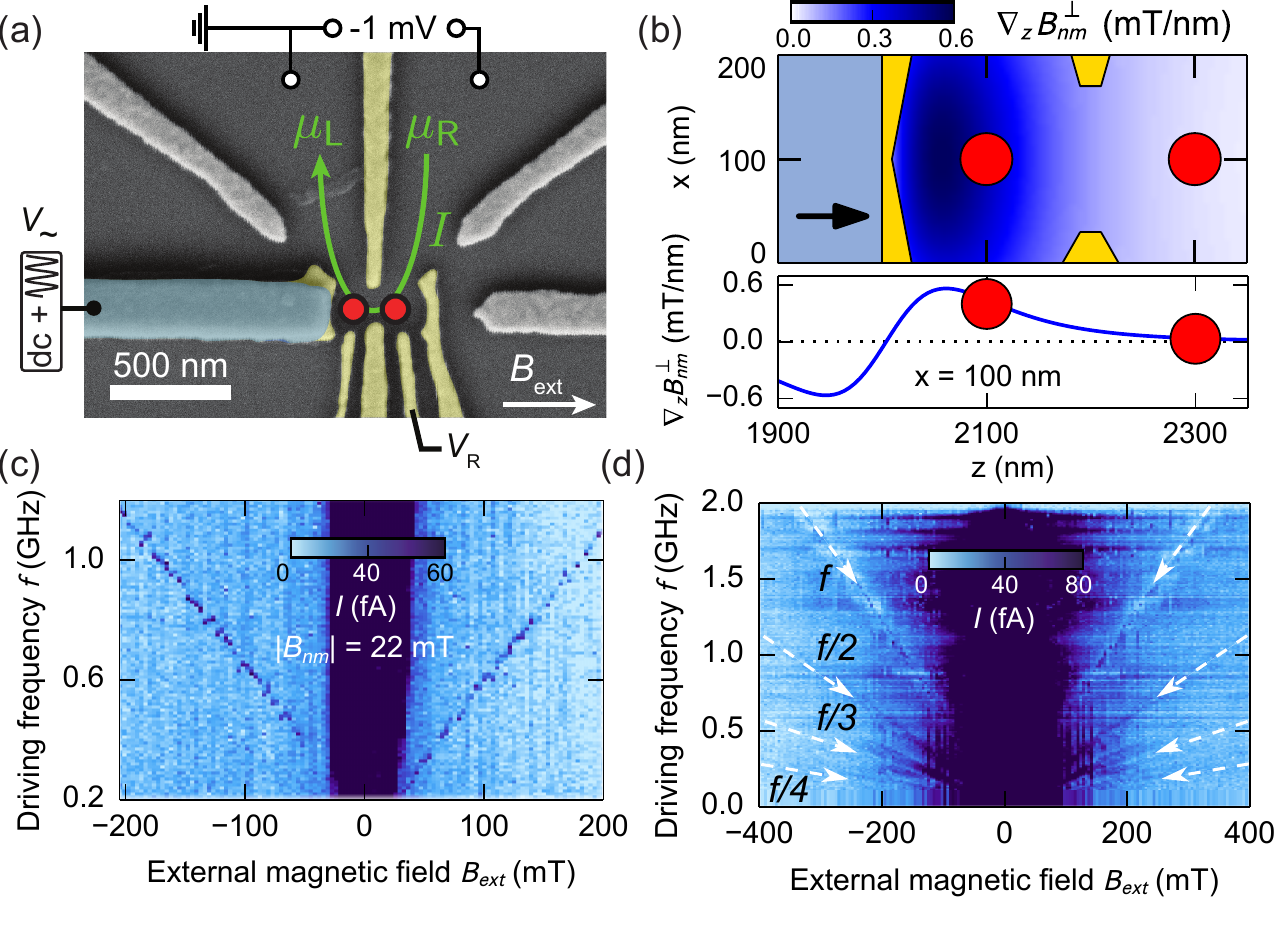}

\caption{Double quantum dot with one instead of two single domain nanomagnets reproduced from Ref.\ \cite{Petersen2013a}. (a) Scanning electron microscope image; color coding analogue to \ref{fig:sample}a. (b) OOMMF simulations of the perpendicular field gradient along \Bext\ aligned with the $z$-axis. This is also the main direction of the rf electron displacement. The lower panel is a horizontal cross section of the color plot at $x=100\,$nm. (c) $I(\Bext, f)$ through the DQD as in Figs. \ref{fig:singleresesr}b, \ref{fig:doubleresesr}a showing a single EDSR resonance at a modulation amplitude of  $\simeq 15\,\text{mV}$. (d) Multiple EDSR resonances at about twice the modulation power.}
\label{fig:single_magnet}
\end{figure}

In the following, we discuss an EDSR measurement of a sample of our previous generation of double QD designs \cite{Petersen2013a}. It contains just one single domain  nanomagnet (see \fig{fig:single_magnet}a) located on top of a gold gate. The voltage on the same gate is modulated to drive EDSR. This design is especially simple as the magnet axis, the rf-modulated gate and the symmetry axis of the QD ($z$-axis) coincide.  It is justified to assume that the gate voltage modulation entails a motion of the QD electrons mostly also along the $z$-axis. Simulations of the nanomagnet shown in \fig{fig:single_magnet}b reveal a sizable magnetic field gradient at the left QD while it almost vanishes at the right QD at a larger distance to the magnet. Our measurements agree with the simulations and reveal an average field of approximately $|\Bm| \simeq 22\,$mT within the left QD and a coercive field of the nanomagnet of $52 \pm 2\,$mT. As in our EDSR experiments, the rf-magnetic field modulation is produced by driving an electron along a time-independent slanting magnetic field, only electrons in the left QD can be manipulated by EDSR. Consequently, in this sample only a single EDSR resonance corresponding to the left QD is expected. This is in contrast to our sample with two nanomagnets, which can be tuned such that a sizable field gradient exists in two QDs giving rise to two EDSR resonances associated with two separate QDs (see \fig{fig:doubleresesr}). As expected, we find only a single EDSR line in the experiment in \fig{fig:single_magnet}c. The $g$-factor is identical to the one of our previous sample, $g=0.36\pm0.01$, while both samples feature similar QDs based on the same wafer. 

In \fig{fig:single_magnet}{d} we present a similar EDSR measurement as in \fig{fig:single_magnet}{c}, but with about twice the modulation power. We observe a transition from a single resonance at $\hbar\omega=g\mB B_\text{ext}$ at small modulation powers in \fig{fig:single_magnet}{c} to multiple resonances at $n\hbar\omega=g\mB B_\text{ext};\, n=1,2,\dots$ for larger modulation powers in \fig{fig:single_magnet}{d}. Such a behavior can be explained in terms of higher order harmonics generation and has also been observed to such a high order in comparable EDSR experiments in an InAs nanowire based double QD \cite{Stehlik2014} and to second order in another GaAs based double QD \cite{Laird2009}. A related example of higher order harmonics generation are Landau-Zener-Stückelberg-Majorana interference oscillations \cite{Forster2014, Rudner2014}. The multiple resonances here (with variable slopes) are fundamentally different from the two resonances (with identical slopes) observed in our previous sample with two nanomagnets. There, they are caused by the different magnetic fields in two QDs and, in contrast to the higher harmonics, observable for low driving powers.

\section{Conclusions}

In summary, we have explored a hybrid nanostructure consisting of a double QD incorporating two single domain nanomagnets with different coercive fields. The magnets' properties agree well with numerical simulations. By sweeping an external field it is possible to reverse the magnet polarizations one-by-one and to directly measure their coercive fields. The magnetization of each nanomagnet switches almost instantly at its coercive field as expected for a single domain magnet at very low temperature. Each switching event modifies the local magnetic field distribution and gives rise to a distinct current jump in a dc transport measurement in the Pauli-spin blockade regime. In a radio frequency EDSR experiment the switching of a magnet generates a shift of the resonance frequency. Compared to the larger multiple domain magnets our single domain magnets generate sizable field gradients even at zero external field and therefore allow experiments in a regime where the relative field difference between adjacent QDs is stronger than the average field. In contrast to magnets with multiple domains, the single domain nature guarantees a stable field distribution over a large range of external field values. The disadvantage of a somewhat smaller field gradient due to the smaller size of the magnet can be compensated by using multiple single domain nanomagnets. The combination of several magnets provides a high\scrap{er} degree of control of the overall field distribution as the coercive fields of each magnet can be predetermined by design.
In summary, coupled QDs including multiple single domain nanomagnets represent a promising approach for future spin qubit circuits desired for quantum information or related spintronics applications.         

We are grateful for financial support from the DFG via SFB-631 and the Cluster of Excellence \emph{Nanosystems Initiative Munich}. S.\,L.\ acknowledges support via a Heisenberg fellowship of the DFG.

\section{Appendix}

\appendix{Methods:}
Magnetic field simulations were carried out with the 3D solver oxsii 1.2a5 of the OOMMF toolkit \cite{Donahue1999}. A simulation gridsize of $5\,$nm (in plane) and $2.5\,$nm (out of plane) was chosen, comparable to the magnetic exchange length of $\sim3.5\,$nm in cobalt. The exchange stiffness of cobalt, $30 \times 10 ^{-12}\,$J/m, as well as the saturation magnetization, $1.4 \times 10^6\,$A/m, enter the simulation as external parameters. We included a reasonable $10^\circ$ correction between the 2DES-plane and the direction of \Bext~which provides the best fit to the measured sample characteristics.


\bibliography{ESR}

\end{document}